\documentclass[sigconf]{acmart}

\usepackage{hyperref}
\usepackage{csquotes}
\usepackage{enumitem}
\usepackage{booktabs}
\usepackage{tabularx}
\usepackage{makecell}
\usepackage[most]{tcolorbox}
\usepackage{graphicx}
\usepackage{natbib}

\usepackage{tikz}
\usetikzlibrary{shapes.geometric, arrows, fit}
\tcbset{
  interviewquote/.style={
    enhanced,
    breakable,
    colback=orange!5,
    colframe=orange!20,
    left=1em,
    right=1em,
    top=0.5em,
    bottom=0.5em,
    boxrule=0pt,
    borderline west={2pt}{0pt}{orange!60},
    sharp corners,
  }
}
\newcommand{\interviewquote}[2]{%
  \begin{tcolorbox}[interviewquote]
    \emph{``#1''} #2
  \end{tcolorbox}
}

\setcopyright{rightsretained}
\copyrightyear{2025}
\acmYear{2025}
\acmDOI{XXXXXXX.XXXXXXX}
\acmConference[Anonymous Conf.]{Anonymous Conference}{2026}{World}

\author{Shalini Chakraborty}
\affiliation{%
  \institution{University of Bayreuth}
  \city{Bayreuth}
  \country{Germany}
}
\email{s.chakraborty@uni-bayreuth.de}
\orcid{0000-0002-9466-3766}

\author{Sebastian Baltes}
\affiliation{%
  \institution{Heidelberg University}
  \city{Heidelberg}
  \country{Germany}
}
\email{sebastian.baltes@uni-heidelberg.de}
\orcid{0000-0002-2442-7522}

\begin{document}

\title{A Multifaceted View on Discrimination in Software~Development~Careers}
\begin{abstract}
Conversations around diversity and inclusion in software engineering often focus on gender and racial disparities. However, the \emph{State of Devs 2025} survey with 8{,}717 participants revealed that other forms of discrimination are similarly prevalent but receive considerably less attention.
This includes discrimination based on age, political perspective, disabilities, or cognitive differences such as neurodivergence.
We conducted a secondary analysis of 800 open-ended survey responses to examine patterns of perceived discrimination, as well as related challenges and negative impacts.
Our study covers multiple identity facets, including age, gender, race, and disability.
We found that age- and gender-related discrimination was the most frequently reported workplace issue, but discrimination based on political and religious views emerged as further notable concerns.
Most of the participants who identified as female cited gender as the primary source of discrimination, often accompanied by intersectional factors such as race, political views, age, or sexual orientation.
Discrimination related to caregiving responsibilities was reported by all gender identities. 
Regarding the negative impacts of workplace issues, many participants described modifying their appearance or behavior in response to gender biases. Gender also appeared to influence broader career challenges, as women and non-binary respondents reported experiencing almost all workplace issues at higher rates, particularly discrimination (35\%) and mental health challenges (62\%).
Our goal is to raise awareness in the research community that discrimination in software development is multifaceted, and to encourage researchers to select and assess relevant facets beyond age and gender when designing software engineering studies.
\end{abstract}



\maketitle

\section{Introduction}
\label{sec:introduction}

Discrimination in the tech industry, particularly in software engineering (SE), remains a persistent issue despite decades of efforts to promote diversity and inclusion (D\&I). Much of the previous research has focused on gender and racial biases in hiring, promotion, and workplace dynamics \cite{vasilescu2015gender, rodriguez2021perceived, terrell2017gender}. However, comparatively fewer studies investigate broader forms of inequity, such as ageism~\cite{DBLP:journals/software/BaltesPS20, DBLP:conf/icse/BreukelenBBS23}, neurodiversity-related bias~\cite{morris2015understanding}, or discrimination due to sexual orientation~\cite{boman2024breaking}, which also shape career progression, workplace experiences, and well-being. 
Representation gaps remain stark across the industry. From 2014 to 2022, the US tech workforce showed minimal progress: in 2022, only 23\% of tech workers were women, despite women constituting nearly half of the general workforce. 
Similarly, Black and Hispanic workers accounted for just 7.4\% and 10\% of the sector, respectively, compared to 11.6\% and 18.7\% across all industries~\cite{reuters2024lackdiversity}. These persistent disparities not only hinder individual advancement but also limit the diversity of perspectives crucial for innovation in software development. 

Beyond gender and race, \emph{ageism} has emerged as another critical but underexplored issue in SE. Older professionals often face biases regarding adaptability, technical competence, or cultural fit~\cite{DBLP:journals/software/BaltesPS20, DBLP:conf/icse/BreukelenBBS23}. At the same time, younger workers may face stereotypes about immaturity or lack of experience~\cite{schloegel2018age}, indicating that age-related discrimination can affect multiple groups in different ways.  Older professionals, despite their experience, remain underrepresented in the tech workforce: in 2022, only 52\% of US tech workers were over 40, down from 55.9\% in 2014~\cite{linkedin2022ageism}. Surveys report that up to 90\% of US workers over 40 have felt age discrimination at work, and 76\% of respondents in a global survey perceived ageism as real within the industry. Formal age-discrimination charges account for nearly one in five workplace complaints in tech, a higher rate than in many other industries. 

\emph{Neurodiversity} is another dimension of discrimination in SE. Neurodivergent individuals, including those with ADHD, Autism Spectrum Disorder (ASD), dyslexia, or dyspraxia, often have valuable cognitive strengths such as enhanced pattern recognition capabilities or attention to detail \cite{gama2025socio}. However, they continue to face systemic barriers in recruitment, communication, and workplace support \cite{verma2025differences}. A recent systematic review identified 18 success factors, ranging from inclusive educational pathways to adaptive workplace technologies, that can improve autistic individuals' career outcomes in Information and Communication Technology (ICT) \cite{sarker2025inclusive}. Nonetheless, many neurodivergent professionals report limited advancement opportunities, high burnout, and a reluctance to disclose their identity due to fear of negative consequences~\cite{codefirstgirls2024neuro}. 

LGBTQ\texttt{+} professionals in tech face layered challenges, from microaggressions to systemic exclusion. Over one-third of LGBTQ\texttt{+} tech employees report wage disparities, some earning between £10,000 and £14,999 less than their heterosexual counterparts for the same roles~\cite{McDonald2019wages}. Almost 40\% have witnessed homophobic harassment in major tech firms such as Facebook and Oracle~\cite{business2019discrimination}. In science, technology, engineering, and mathematics (STEM) broadly, LGBTQ\texttt{+} individuals report higher rates of exclusion; 32.9\% experience social marginalization, and they face harassment and health difficulties more frequently than their non-LGBTQ peers~\cite{businessinsider2019lgbtq}. In agile software teams, LGBTQ\texttt{+} professionals have described feelings of invisibility, prejudice, and discrimination---although inclusive environments can improve team participation and developer experience~\cite{wassouf2025developer}. However, these dimensions of discrimination do not occur in isolation. Instead, they often intersect and compound the disadvantage.
For example, a neurodivergent LGBTQ\texttt{+} developer or an older woman developer may face overlapping forms of bias, increasing barriers to inclusion, satisfaction, and retention. Despite growing awareness of intersectionality, SE research has rarely addressed discrimination through a holistic, multifaceted lens.

To address this gap, we performed a secondary analysis of the 2025 \emph{State of Devs} survey~\cite{stateofdevs2025}, which includes responses from 8{,}717 developers worldwide. We followed Altman's~\cite{altman2011discrimination} definition of discrimination, \emph{``discrimination consists of acts, practices, or policies that impose a relative disadvantage on persons based on their membership in a salient social group"} and analyzed 800 open-ended responses focusing on workplace discrimination, related challenges, and perceived negative impacts. Our findings reveal that discrimination extends far beyond gender and race, encompassing issues related to age, political and religious views, disability, and caregiving responsibilities. Furthermore, gender remains a pervasive factor influencing workplace experiences: women and non-binary respondents reported higher rates of discrimination (35\%) and mental health challenges (62\%) than other groups.

Through this work, our goal is to raise awareness within the software engineering research community that discrimination in development careers is multifaceted. 

\begin{figure}[tb]
    \centering
    \includegraphics[width=0.8\linewidth, trim=0 205pt 0 5pt,clip]{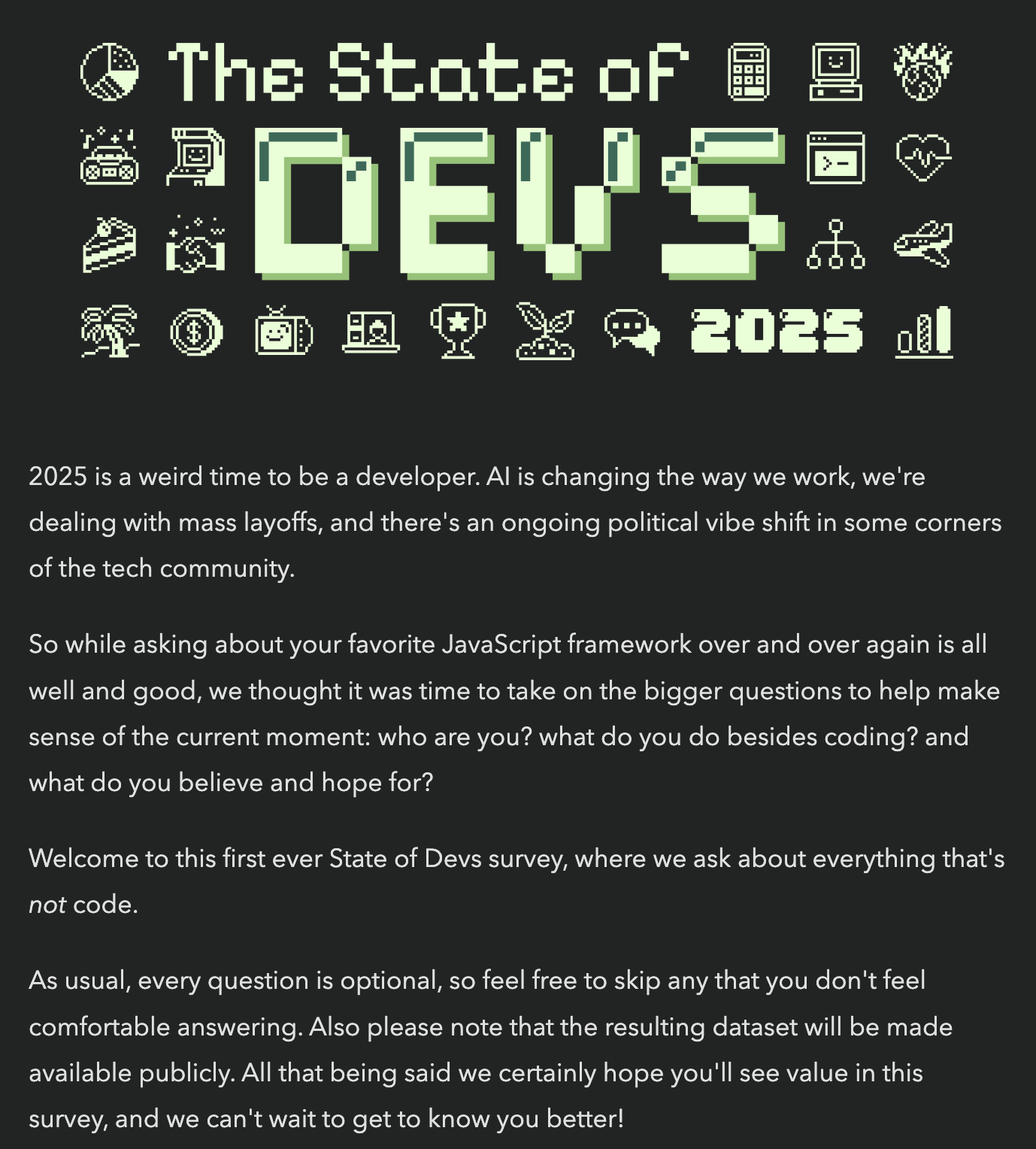}
    \caption{Start page of the \emph{State of Devs 2025} survey~\cite{stateOfDevsPage2025}.}
    \Description{Screenshot showing the start page of the State of Devs 2025 survey.}
    \label{fig:survey-start-page}
\end{figure}

\section{Related Work and Background}
\label{sec:background}
Discrimination in the workplace has long been studied in many domains. It is an issue across industries, influencing hiring, promotion, pay equity, and employee well-being~\cite{dobbin2022getting, reskin2000proximate, pager2009discrimination}.  
Research in organizational psychology and sociology has shown that both overt and subtle forms of bias such as microaggressions and stereotype-based evaluations, negatively affect productivity, job satisfaction, and retention~\cite{hebl2020modern, sue2007racial, williams2021bias}.

\subsection{Discrimination in Software Engineering}  
Within SE, a growing body of empirical work documents how discrimination and bias manifest in developer communities, organizations, and collaborative platforms.  
For example, the study ``Mind the gap: gender, micro-inequities and barriers in software development'' finds that unconscious biases are widespread and affect task assignment, team inclusion, recognition, and value of contributions in SE contexts~\cite{guzman2024mind}.  
Similarly, analyses of open source software communities show that gendered behavior patterns rather than categorical gender membership alone contribute substantially to women’s disadvantage in code review, project leadership, and long-term participation~\cite{terrell2017gender, vedres2019gendered}.  
Further, a qualitative study at a major technology company examined how teams build and sustain gender-diverse software engineering teams, showing that inclusive practices in recruiting, hiring, and team culture play a critical role~\cite{kohl2022benefits}.  
These studies underscore that discrimination in SE is not only a matter of under-representation, but also concerns how technical work is organized, valued, and recognized.  
Age-related discrimination remains also a concern. For example, \citeauthor{van2023still} highlight the challenges faced by veteran women developers in hiring practices~\cite{van2023still}. Similarly, \citeauthor{liang2024controlled} evaluated the effect of demographic information (gender and age) on the assessment of technical articles in software engineering, finding that raters tend to give more favorable content depth evaluations to younger male authors compared to older male authors, and that male participants complete evaluations faster than female participants~\cite{liang2024controlled}. 
However, most SE research still emphasizes gender, age, or ethnicity, while comparatively fewer works explore other identity facets such as neurodiversity, sexual orientation, caregiving responsibilities, or political views.

\subsection{Spectrum of Identity Dimensions: LGBTQ\texttt{+}, Neurodiversity, and Beyond}
Beyond gender and race, research is increasingly examining other identity dimensions that shape experiences in technical fields.  
For example, a study of engineering education found that LGBTQ+ students experience greater marginalization, devaluation, and health and wellness difficulties compared to their non-LGBTQ+ peers~\cite{cech2018lgbtq}.  
In the realm of neurodiversity, recent work shows that autistic individuals often face barriers in recruitment, workplace support, and social dynamics, despite evidence of valuable cognitive strengths~\cite{sarker2025inclusive}.

These findings suggest that age and gender biases are present even in seemingly objective technical assessments, highlighting the importance of raising awareness before participants can actively report or confront discrimination in software development careers.  

\subsection{Why Discrimination Matters in Software Engineering}  
Discrimination directly affects career progression, retention, well-being, and productivity, which is particularly consequential in an industry already challenged by high attrition and rapidly changing skill demands~\cite{terrell2017gender, ford2016paradise}.  
As SE increasingly relies on collaborative digital platforms, open source communities, remote work, and global teams, subtle biases such as those embedded in tools, algorithms, or task assignments can amplify inequities; for example, gendered behavior patterns in open source projects have been linked to persistent disadvantage~\cite{vedres2019gendered}.  
Emerging identity dimensions, including LGBTQ+ status, neurodiversity, and political views, are often overlooked in studies of workplace discrimination despite their significant impact.  
Our analysis highlights these patterns and serves as a foundation for future research, encouraging the inclusion of a broader set of relevant identity facets beyond the traditional focus on gender, age, and race when designing discrimination studies.

\section{Methodology}

\subsection{Survey Overview}
This study draws on data from the \emph{State of Devs 2025}~\cite{stateofdevs2025} survey, a publicly available annual global survey of software professionals.
Unlike surveys with a more technical focus such as the annual \emph{Stack Overflow Developer Survey}~\cite{stackoverflow2025}, \emph{State of Devs} emphasizes non-technical aspects of developers' lives, including workplace conditions, organizational practices, health and well-being, hobbies, career development, and perceptions of equity and inclusion. 

The 2025 survey is organized into multiple high-level categories. Table~\ref{tab:survey_outline} shows an overview of the survey with seven main categories and their corresponding subcategories. 

\begin{table*}[ht]
  \centering
  \caption{Survey Overview: Categories and Subcategories.}
  \label{tab:survey_outline}
  \small
  \begin{tabularx}{\textwidth}{>{\bfseries}p{2.2cm} X}
    \toprule
    Main Category & Subcategory (or Sub-section) \\ 
    \midrule
    Demographics & Country / Region, Gender, Race / Ethnicity, Relationship Status, Number of Children \\ 
    \midrule
    Career & Years of Experience, Higher Education Degree, Employer Count, Employment Status, Workplace Issues, Discrimination, Negative Impact, Layoffs \\ 
    \midrule
    Workplace & Company Size, Job Title, Yearly Income, Income Evolution, Number of Application, Job Finding, Remote Work, Work Hours, Job Happiness, Workplace Perks, Workplace Difficulties \\ 
    \midrule
    Technology & Mobile OS, Desktop OS, Programming Languages, productivity Apps, Messaging Apps, Social Media, Community Involvement, Community Contributions, Open Source  \\ 
    \midrule
    Health & Age, Disability Status, Health Issues, Sleep Quantity, Sleep Strategies \\ 
    \midrule
    Worldview & Global Issues, Happiness Factors, Relocation, Overall Happiness \\ 
    \midrule
    Hobbies & Hobbies, Physical Activities, Favorite Music, Favorite Movies, Favorite TV Shows, Favorite Video Games, Favorite Game Platforms  \\ 
    \bottomrule
  \end{tabularx}
\end{table*}
 
The sample was collected by the \emph{Devographics} survey team through an open call distributed in online communities, newsletters, and developer networks. Participation was voluntary and anonymous, with informed consent presented before survey submission. Because the data was self-reported, it is subject to common limitations such as self-selection bias and recall bias.
The survey attracted 8,717 participants from a diverse and global sample of developers across North America, Europe, Asia, Latin America, and Africa.
The demographic items in the questionnaire allow stratification by gender, race/ethnicity, sexual orientation, disability/neurodiversity, and age group. These demographic features make the dataset suitable for examining discrimination-related experiences along multiple identity axes.

\subsection{Study Scope and Research Questions}

This paper focuses specifically on the discrimination-related aspects of the survey. Although the \emph{State of Devs 2025} survey covers a broad range of non-technical topics, our analysis isolates items that reflect or imply bias in the workplace. By triangulating quantitative measures with qualitative descriptions of personal experiences, our objective is to develop a holistic view of discrimination in software development careers along multiple facets: gender, race, age, neurodiversity, and LGBTQ\texttt{+}. The paper follows three research questions (RQs):

\begin{description}[style=multiline, labelindent=4mm, leftmargin=12mm]
\item[RQ1] \emph{What is the current state of discrimination across different demographic groups of software developers?}
\item[RQ2] \emph{What are the most common challenges and negative consequences experienced by developers who report discrimination in their careers?}
\end{description}

\subsection{Data Collection}

The complete dataset is publicly available through the \emph{Devographics API}~\cite{stateofdevs2025}, which provides structured access to quantitative and qualitative survey responses~\cite{stateOfDevsPage2025}. 
For the purpose of this paper, we extracted a subset of data focusing on workplace experiences and issues related to discrimination, bias, and inclusion. Specifically, we filtered the dataset to include responses to the following three open-ended questions:
\begin{description}[style=multiline, labelindent=4mm, leftmargin=10mm]
\item[Q1] \emph{``Which of these workplace issues have you experienced throughout your professional career?''}
\item[Q2] \emph{``Have you ever experienced discrimination in the workplace based on any of the following factors?''}
\item[Q3] \emph{``Have you taken any of the following actions in response to workplace issues?''}
\end{description}

These questions were selected because they directly or indirectly address aspects of workplace treatment.
In total, we qualitatively analyzed 800 open responses.
Responses were included if they described personal or observed instances of discrimination, unfair treatment, or workplace inequity.  
The final dataset used for analysis therefore represents a focused view of participants' experiences with discrimination and related workplace challenges in software development career. We publish the dataset as part of our supplementary material~\cite{anonymous2025multifaceted}.

\subsection{Data Analysis}
We adopted a mixed methods approach that combines quantitative and qualitative analyses to provide a comprehensive view of discrimination experiences in software development careers. 

\subsubsection{Quantitative Analysis}
To address \textbf{RQ1}, we performed descriptive statistical analyses to estimate the multifaceted nature of discrimination across demographic groups, including gender, age, race or ethnicity, neurodiversity, and LGBTQ\texttt{+} identity (see Section~\ref{sec:quantitaive}). Based on \textbf{Q1} and \textbf{Q2}, we performed cross-tabulations and Chi-square tests to explore associations between demographic variables and self-reported discrimination experiences. This analysis provided an overview of the prevalence and distribution of perceived discrimination among different identity groups.

\subsubsection{Qualitative Analysis}
To answer \textbf{RQ2}, we performed a qualitative analysis of open-ended survey responses from participants who reported experiences of discrimination (see Section~\ref{sec:qualitative}). Using descriptive coding, we identified recurring topics related to discriminatory experiences, challenges, and negative impacts. We iteratively refined the codes and grouped them into higher-level themes that represent how developers perceive and navigate discriminatory environments. The first author conducted the initial round of coding, followed by an independent review by the second author. The two researchers then compared and discussed their findings to reach a consensus. 

We began our qualitative analysis with responses to \textbf{Q2} (\emph{``Have you ever experienced discrimination in the workplace based on any of the following factors?''}). Using the API provided by the survey organizers, we retrieved 241 open-ended responses for this question. The first author categorized these responses according to the discrimination factors defined by the survey (e.g., gender, age, disability, race, political views, sexual orientation). During this process, three additional categories were introduced: \emph{na} (responses indicating that no discrimination was experienced), \emph{observed} (responses describing discrimination observed in others) and \emph{other} (responses referencing general unfair treatment not covered by existing categories). In cases where a response referenced multiple factors, we designated a \emph{primary discrimination factor} and recorded additional references as \emph{secondary discrimination factors}.

For \textbf{Q1} (\emph{``Which of these workplace issues have you experienced throughout your professional career?''}), we analyzed 447 responses and retrieved 95 that explicitly mentioned ``discrimination'' as a workplace issue. Each response was coded along three dimensions: (1) whether it addressed discrimination, (2) the \emph{primary discrimination factor}, and (3) any \emph{secondary discrimination factor}. The discrimination categories used were consistent with those from \textbf{Q2}, including the additional \emph{na}, \emph{observed}, and \emph{other} categories. 

Finally, for \textbf{Q3} (\emph{``Have you taken any of the following actions in response to workplace issues?''}), we retrieved 112 responses describing negative workplace impacts. Because these comments referenced a variety of workplace issues and not exclusively discrimination, we manually reviewed and classified them into three categories: (1) whether the negative impact was discrimination-related, (2) associated discrimination factors, and (3) described outcomes or consequences. 


\section{Results}
\subsection{RQ1: Current state of discrimination}
\label{sec:quantitaive}
To examine the current state of discrimination, we leveraged two survey questions (\textbf{Q1} and \textbf{Q2}) from the career section of the questionnaire. First, respondents were asked: \textit{Which of these workplace issues have you experienced throughout your professional career?}. 8,516 participants answered this question. Among multiple options was ``Discrimination”. 13\% voted it as a workplace issue, which is 1101 survey participants. Gender does seem to have an impact on career issues, with women and non-binary respondents experiencing every issue at a higher rate, especially when it comes to discrimination (35\%) and mental health issues (62\%).

Second, participants were asked specifically if they had ever experienced discrimination in the workplace based on specific factors such as age, gender, race, disabilities, or sexual orientation. The discrimination factors item had responses from 7,467 participants (~86 \% response rate) for that question; participants could select multiple factors.
Among all respondents (7467), \emph{age} is the most frequently cited basis of discrimination (12\% or 896 respondents), followed by \emph{gender / gender identity} (11\% or 835), \emph{disabilities} (6\% or 470), \emph{Political Beliefs} (5\% or 402) and \emph{race / ethnicity} ( 5\% or 368).

Table~\ref{tab:discrimination_factors} shows the different discriminations cited and the percentage of participants who voted for each discrimination. 

\begin{table}[tb]
\centering
\small
\caption{Reported discrimination factors (n = 7,467).}
\label{tab:discrimination_factors}
\begin{tabular}{lrrr}
\hline
\textbf{Discrimination Factor} & \textbf{Count} & \textbf{Percentage}\\
\hline
Age & 896 & 12.0\% \\
Gender & 835 & 11.18\% \\
Disability & 470 & 6.29\% \\
Political Views & 402 & 5.38\% \\
Race & 368 & 4.93\% \\
Caregiving Status & 293 & 3.92\% \\
Sexual Orientation & 216 & 2.89\% \\
Religion & 200 & 2.68\% \\
Country / Citizenship & 48 & 0.64\% \\
Appearance / Body & 10 & 0.13\% \\
Not Applicable & 5251 & 70.32\% \\
Other Answers & 156 & 2.09\% \\
\hline
\end{tabular}
\end{table}

Table~\ref{tab:age_discrimination} illustrates the proportion of respondents in each age group who reported experiencing ``Discrimination” as one of their workplace issues. The trend reveals a clear upward gradient.

\begin{table*}[tb]
\centering
\small
\caption{Discrimination reported by age group (n = 7,467).}
\label{tab:age_discrimination}
\begin{tabular}{c*{9}{r}}
\toprule
\textbf{Age Group} & \textbf{Age} & \textbf{Gender} & \textbf{Disability} & \textbf{Political} & \textbf{Race} & \textbf{Caregiving} & \textbf{Sexual} & \textbf{Religion} & \textbf{Other/} \\
& & & & \textbf{Views} & & \textbf{Orientation} & & & \textbf{No Answer} \\
\midrule
$<$20 & 0 & 0 & 0 & 0 & 0 & 0 & 0 & 0 & 62 \\
20-29 & 153 & 134 & 66 & 72 & 41 & 0 & 0 & 0 & 1307 \\
30-39 & 274 & 350 & 181 & 148 & 162 & 0 & 0 & 0 & 2582 \\
40-49 & 176 & 189 & 99 & 73 & 0 & 98 & 0 & 0 & 1416 \\
50-59 & 89 & 30 & 32 & 28 & 0 & 22 & 0 & 0 & 342 \\
$\ge$60 & 43 & 11 & 10 & 11 & 0 & 0 & 0 & 0 & 33 \\
\bottomrule
\end{tabular}
\end{table*}

The table shows that respondents in older age brackets report discrimination at higher rates. For example, in the 50–59 age group, 89 participants cited age discrimination and 30 cited gender, while in the $\ge$60 group, 43 reported age and 11 reported gender discrimination. In contrast, younger groups such as 20–29 reported 153 cases of age discrimination and 134 of gender, and 30–39 reported 274 and 350, respectively.
Although the absolute numbers are higher in younger groups due to larger sample sizes, the proportion relative to the total responses per age bracket indicates that older respondents experience or perceive discrimination more frequently. For example, age discrimination constitutes 16. 4\% of the reported issues in the 50–59 group and 27.0\% in the $\ge$60 group, compared to 8.6\% in 20–29 and 7.4\% in 30–39.

We performed a Chi-square test of independence to examine the association between age group and reporting discrimination. The test revealed a significant relationship, $\chi^2(5, N=7467) = 112.4, p < 0.001$, indicating that the likelihood of reporting discrimination varies across age groups.
Post hoc pairwise comparisons using odds ratios with Bonferroni-adjusted p-values showed that respondents aged $\ge$60 were significantly more likely to report discrimination than the 20-29 baseline group (OR = 3.25, 95\% CI = 2.10-5.03, $p < 0.01$). Similarly, the 50--59 age group had elevated odds compared to the baseline (OR = 2.15, 95\% CI = 1.65-2.80, $p < 0.01$). Younger groups (30-39 and 40-49) also reported higher discrimination than the 20-29 group, but with smaller effect sizes (ORs = 1.10-1.45). 
This gradient suggests that discrimination is meaningfully associated with age: While younger developers do experience discrimination, the likelihood of reporting it increases with age, emphasizing that age-related workplace bias remains a persistent issue in software engineering.


Table~\ref{tab:gender_discrimination} shows the discrimination reported in the workplace between men and non-men respondents. Gender-based discrimination is the most frequently reported issue for non-men respondents, affecting 47\% of the group, while only 3.5\% of men report it. Age, disability, and caregiving-related discrimination are also reported more often by non-men respondents, although age discrimination is notable among men as well. Sexual orientation is another area with higher prevalence among non-men respondents (7.7\% vs. 1.9\%). In contrast, men more frequently report political-view discrimination and, to a lesser extent, race or religion. The ``Other / No Answer'' category indicates that a substantial portion of respondents did not select any predefined discrimination types, particularly among men. Overall, the table highlights stark differences in the types and frequencies of discrimination experienced across gender identities, underscoring the importance of examining multiple dimensions of workplace bias in software engineering.
While for participants identifying as men voted ``age'' to be the their number one discrimination issue (575 among 5280), among non-men it is gender or gender identity (558 among 1181). 

\begin{table}[tb]
\centering
\small
\caption{Reported workplace discrimination by gender identity; percentages are relative to each group (n = 5,280).}
\label{tab:gender_discrimination}
\renewcommand{\arraystretch}{1.1}
\begin{tabular}{lrr}
\toprule
\textbf{Discrimination Type} & \textbf{Men (\%)} & \textbf{Non-Men (\%)} \\
\midrule
Age & 10.9 & 16.9 \\
Gender & 3.5 & 47.3 \\
Disability & 4.7 & 14.7 \\
Political Views & 5.3 & 5.7 \\
Race & 4.4 & 6.2 \\
Caregiving Status & 3.2 & 7.0 \\
Sexual Orientation & 1.9 & 7.7 \\
Religion & 2.6 & 1.9 \\
Country / Citizenship & 0.7 & - \\
Other / No Answer & 94.4 & 54.6 \\
\bottomrule
\end{tabular}
\end{table}

Table~\ref{tab:race_discrimination} summarizes the distribution of reported workplace discrimination across racial and ethnic groups. White respondents most frequently reported discrimination based on age, gender, disability, and political views, whereas non-White respondents reported race-related discrimination as the most common form, followed by gender and age. The table highlights the distinct patterns of discrimination experienced by different racial/ethnic groups, showing that while some forms of discrimination are widespread across all groups (e.g., gender and age), others (such as race) disproportionately affect non-White developers. Caregiving responsibilities and sexual orientation are reported less frequently among non-white respondents. 

\begin{table}[tb]
\centering
\small
\caption{Reported workplace discrimination by racial/ethnic group; percentages are relative to each group (n = 4,710).}
\label{tab:race_discrimination}
\begin{tabular}{lrr}
\toprule
\textbf{Discrimination Type} & \textbf{White (\%)} & \textbf{Non-White (\%)} \\
\midrule
Age & 11.9 & 11.7 \\
Gender & 11.6 & 11.3 \\
Disability & 7.3 & 4.2 \\
Political Views & 5.3 & 5.6 \\
Race & 2.9 & 11.7 \\
Caregiving Status & 4.3 & 2.7 \\
Sexual Orientation & 3.2 & 2.4 \\
Religion & 2.3 & 3.0 \\
Country / Citizenship & 0.8 & - \\
Other / No Answer & 70.7 & 69.1 \\
\bottomrule
\end{tabular}
\end{table}

\subsection{RQ2: Most common challenges and negative consequences of discrimination}
\label{sec:qualitative}
To complement our quantitative findings and address \textbf{RQ2}, we conducted a qualitative content analysis of three open-ended survey questions (\textbf{Q1}, \textbf{Q2}, and \textbf{Q3}). We present our findings in the following subsections:

\subsubsection{Discrimination as a Workplace Issue and Its Effects}

For \textbf{Q1}, \emph{``Which of these workplace issues have you experienced throughout your professional career?''}, 95 of 447 participants explicitly mentioned \emph{discrimination} as a workplace issue. Among these, the majority (36 responses) referred to gender-related discrimination, primarily directed toward women. Respondents described experiences of exclusion, microaggressions, and unequal treatment:

\interviewquote{I've only ever worked on dev teams where I am the only woman, so plenty of microaggressions and being treated differently from my team.}

\interviewquote{I'm the only female developer in my company and the only one who hasn't received a pay rise offer after a year at the company.}

Transphobia was another recurring concern, often intertwined with sexism and social exclusion:

\interviewquote{I've had to deal with a lot of transphobia as a trans man, or otherwise being excluded from events for being trans.}

\interviewquote{Mostly transphobia and sexism, including ``funny'' comments, inappropriate questions, and physical contact, especially at events and conferences. People belittle you, don't believe you're a developer, and still try to chat you up.}

Several participants also mentioned discrimination related to race or nationality. For example, respondents reported workplace jokes about ethnicity or differential treatment based on visa dependency:

\interviewquote{Discrimination was more about making jokes behind my back about my ethnicity.}

\interviewquote{I've witnessed managers bullying subordinates who were tethered to the job by visas, and personally experienced a co-worker actively working to undermine my team.}

Because participants were not provided with a predefined list of discrimination factors before answering this question, several responses were categorized as \textbf{other}, referring to more general or ambiguous forms of unfair treatment:

\interviewquote{Discrimination was only trivial. Stress and anxiety are part of work—depending on an individual's personality. Even a normal working week isn't really a ``balance.''}

Although gender-related issues were the most frequently mentioned, some respondents highlighted discrimination based on sexual orientation:

\interviewquote{It's a rough field for anyone who is not a cis-male out there.}

Interestingly, some responses framed discrimination not around identity but around job roles, particularly the perceived undervaluation of certain developer positions or the impact of AI on job security:

\interviewquote{Discrimination against my input about backend problems faced by the frontend because I am a frontend dev, even though I have backend experience.}

\interviewquote{Common theme seems to be making me feel like I suck as a developer so they pay me less, or now with AI, making me feel like I can easily be replaced—so no need to pay me.}

Participants frequently described mental strain, burnout, and job insecurity as consequences of these experiences:

\interviewquote{Burnout, stress, work-life balance, insufficient wages… those are basically considered normal in early-stage startups, and this is where I usually work. As for harassment and discrimination, it's common everywhere.}

Overall, responses to \textbf{Q1} depict discrimination as a multifaceted and widespread workplace issue, most prominently related to gender, but also shaped by trans identity, race, sexual orientation and professional roles.

\subsubsection{Different Factors of Discrimination}

In Section~\ref{sec:quantitaive}, we observed that \emph{age} was the most frequently cited discrimination factor. The open-ended responses to \textbf{Q2} provide a more nuanced and detailed picture of how discrimination manifests across multiple identity dimensions. Participants described experiences related to age, gender, race, political beliefs, disability, caregiving responsibilities, and other factors.

\paragraph{Age Discrimination.} 
We identified 26 specific accounts of age-related discrimination, highlighting both ends of the age spectrum. Some respondents reported being perceived as too young or inexperienced despite their competence:
\interviewquote{Being seen as too young to know enough when I first started, even though I knew more about the actual work involved in the company products than my startup bosses was painful.}

Others described ageism directed toward older employees:
\interviewquote{I'm 37, people remind me of that all the time.}

\paragraph{Gender Discrimination.} 
Gender bias appeared as the most prominent theme across responses. Women, in particular, described facing stereotyping, condescension, and professional undermining:
\interviewquote{Yep, being a woman in a man's world. And being called 'wet behind the ears' in professional settings where I was the team lead, just to put me down.}

Several respondents described intersecting experiences where gender discrimination overlapped with other factors such as age:
\interviewquote{I am a woman and often get told I look much younger than my age. I feel like because of a combination of these two factors, I sometimes get condescended to and underestimated.}

\paragraph{Racial and Ethnic Discrimination.}
Racial bias emerged as another recurring factor, often expressed through microaggressions or overt exclusion:
\interviewquote{Making jokes behind my back when I arrived; I was the first non-white person to join the company. Someone made a joke like, ``I never thought that you would recruit those kinds of persons one day.'' I heard about it years later.}

\interviewquote{I have been falsely accused of believing things I do not believe and doing things I did not do on the basis of my race. These were not misinterpretations, but complete fabrications.}

\paragraph{Political Beliefs.} 
We analyzed 20 responses describing discrimination based on political views. Respondents mentioned being treated differently for holding conservative or libertarian perspectives:
\interviewquote{Not belonging to the far-left seems to be a problem in many software workplaces.}

\interviewquote{People have treated me differently for having more conservative-libertarian or non-liberal views.}

Some narratives reflected tensions between political neutrality and diversity initiatives within organizations:
\interviewquote{The company introduced a policy of not talking about politics at work and basically shut down all of its DEI initiatives, making the associated Slack channels a very unsafe space.}

Others criticized poorly implemented diversity policies that unintentionally created exclusion:
\interviewquote{The biggest discrimination I've seen came from poorly implemented DEI policies. I witnessed two discriminatory policies before leaving: one that denied work to established male technicians in favor of more diverse hires with empirically worse track records, and another where a manager said, ``The last four hires were all men. Stop bringing me male applicants. The next technical hire should be a woman.''}

\paragraph{Disability and Neurodiversity.} 
Discrimination based on disability was also mentioned, often associated with a lack of accommodation:
\interviewquote{I was laid off from my last non-developer role because I was disabled and they did not want to make any accommodations for me. I'd been there for 18 years.}

Although the survey grouped neurodiversity under disability, we argue (as discussed in Section~\ref{sec:discussion}) that neurodivergence should be treated as a distinct facet of human diversity rather than a disability. Several respondents reflected on challenges linked to neurodivergent traits:
\interviewquote{I have ADHD and a non-linear thinking pattern. I saw on a review once that I needed to take a more ``cognitive'' approach to my tasks, which left me wondering what that meant, since no one directly approached me or clarified what I needed to do differently.}

\interviewquote{I also was denied when I requested to implement an email-based ticket tracking system for me and my coworkers to use which would have dramatically helped with my ADHD.}

\paragraph{Caregiving Responsibilities.} 
Caregiving emerged as a source of bias across all gender identities:
\interviewquote{Overlooked for promotion because I have a child. The manager just assumed I didn't want it and never discussed it with me.}

\interviewquote{As a white guy taking care of my kids alone, I was ignored and snubbed.}

\paragraph{Other Factors.} 
Some respondents mentioned additional bases of discrimination, such as nationality, sexual orientation, religion, and physical appearance.  
For example, some described language-related exclusion:
\interviewquote{I feel like I am less taken into account for not having the same native language as the rest of the team.}

Others shared experiences related to sexual orientation:
\interviewquote{I've often found it difficult to navigate being open about being gay because you can never tell what someone thinks. With the rise of remote working, this has become even worse—being social feels like a waste of time, so the topic hardly comes up naturally.}

Religious bias also appeared in some accounts:
\interviewquote{One of my bosses was weird about me going to church occasionally. Ageism I experienced going into leadership roles where people perceived me as too young to be in a management role.}

Finally, some respondents mentioned bias based on physical appearance or conformity to workplace norms:
\interviewquote{In one workplace I was told I'd never be taken seriously as someone who works part time. I was told I didn't ``look the part.'' I was told to ``lose the dyke haircut.''}

Overall, these qualitative responses reveal that discrimination in software development is highly multifaceted. Although gender and age remain dominant themes, participants also experience exclusion based on political identity, disability, caregiving status, and appearance, illustrating how important it is to identity all discrimination intersect to shape workplace inequities.

\subsubsection{Not Felt but Observed}
We observed many responses that indicated that the participants had not personally experienced discrimination. However, even when not felt directly, their comments often acknowledged the presence of discrimination toward others in the workplace. Notably, most of these responses came from participants who identified themselves as ``straight/cisgender white men.'' 
 
\interviewquote{I'm a straight, white, middle class man so I only see it rather than experience it}

\interviewquote{I have not been discriminated against in the workplace, but as a gender-conforming, straight, white man, that statement shouldn't be taken to mean "discrimination hasn't taken place."}

\interviewquote{I'm blessed with being a white CIS male. I've witnessed discrimination, but never been the target.}

\subsubsection{Negative Impacts of Discrimination}

To explore the negative consequences of workplace issues, we analyzed 112 open-ended responses describing perceived negative impacts. Although these accounts covered a wide range of workplace challenges and were not all directly related to discrimination, we systematically identified and categorized those that were clearly related to discriminatory experiences.

One of the most frequently reported coping strategies involved changing personal behavior or self-presentation in response to workplace discrimination. For example, several neurodivergent respondents described altering their communication style to mask traits associated with autism or ADHD:
\interviewquote{I used to mask my speech (autism) when I was fresher in my job.}

Changes in appearance or self-expression were also common, especially among those facing gender- or age-based discrimination:
\interviewquote{I have reported inappropriate comments from certain male colleagues to HR. Over the years I have lowered the pitch of my voice to sound more authoritative in order to be taken more seriously.}

\interviewquote{Changed the way you dress: I cut my hair and dress more masculine.}

\interviewquote{I dress a lot more professionally than the men in my team and have to be a lot more confident in my statements to get the same result.}

\interviewquote{Dyed my hair so I'd appear younger.}

\interviewquote{To minimize unwelcome comments about my appearance, I wear loose or baggy clothing in muted colors and try to keep my appearance as unremarkable as possible. (I am a woman in a male-dominant industry—remarks always happen, even with this strategy.)}

In addition to these behavioral adjustments, many participants described mental health challenges such as stress, anxiety, and burnout:
\interviewquote{Started some ADHD and anxiety meds, which helped in the high-stress situations.}

A recurring consequence of these experiences was resignation or career change. Interestingly, participants often did not attribute their decision to leave directly to discrimination, but rather to the cumulative effects of related factors, such as toxic workplace culture, poor management, or prolonged burnout:
\interviewquote{I wish I could have taken the options above, however I was never given the time or the opportunity to seek help. The times that I asked management for assistance normally led me back to square one. It lingers with me to this day where when I asked my Manager how I can improve their response was just simple "Change your mindset".}

These findings suggest that discrimination and its indirect consequences, such as emotional exhaustion and workplace dissatisfaction, can have profound and long-lasting effects on the well-being of developers and their career trajectories. They also highlight how subtle, persistent forms of bias can drive attrition without being explicitly recognized as discrimination.

\section{Discussion}
\label{sec:discussion}

Our analysis of the \emph{State of Devs 2025} survey provides a multifaceted view of how discrimination manifests itself in software development careers. The results show that while diversity and inclusion initiatives have made progress, subtle and systemic forms of bias persist. In this section, we discuss five key insights that emerged from our analysis.

\subsection{Discrimination is Multifaceted and Still Lingering}

The data suggest that discrimination in software development career is far more complex than the traditionally discussed dimensions of gender and race. Participants reported unfair treatment based on age, political beliefs, caregiving responsibilities, disability, and neurodiversity, highlighting that discrimination operates across multiple, intersecting identity factors.  
Although explicit exclusion practices have decreased, subtle microaggressions and implicit biases remain widespread. This aligns with previous work showing that gender bias, ageism, and ableism in software engineering persist in nuanced forms, shaping hiring, promotion, and team interactions~\cite{vasilescu2015gender, DBLP:journals/software/BaltesPS20, wassouf2025developer, van2023still}.  

Our qualitative findings reveal that developers often internalize these pressures, modifying their behavior or self-presentation in an effort to fit perceived professional norms. These adaptations, such as lowering one's voice, changing one's dress, or concealing identity traits, indicate that social barriers remain embedded in professional culture, despite increasing organizational awareness of equity and inclusion.

\subsection{Neurodiversity Is Not a Disability}


One notable issue in the dataset is the treatment of \emph{neurodiversity} as a subset of disability. Although neurodivergence and disability can overlap, the relationship is not binary. Conditions such as autism are officially recognized as developmental disabilities by organizations such as the \emph{U.S. Centers for Disease Control and Prevention} (CDC)~\cite{cdc_autism_2025}, yet they also exist on a broad spectrum of cognitive variation. The threshold at which a neurodivergent trait is considered a disability is context-dependent and is shaped by environmental barriers and social attitudes. 
Several survey respondents expressed discomfort with labeling neurodivergence as a disability, noting that such a framing may obscure individual strengths and reinforce deficit-oriented perspectives:
\interviewquote{Labeling neurodivergence as a disability is not very neuroaffirming.}

Consistent with previous research, respondents described masking behaviors, such as suppressing neurodivergent communication styles, to appear more ``professional.'' This aligns with broader findings that neurodivergent developers frequently adjust their communication or work habits to avoid stigma~\cite{gama2025socio, verma2025differences}. However, these acts of self-censorship can contribute to burnout and reduced psychological safety. 
Recognizing neurodiversity as a distinct identity rather than a disability would allow organizations to focus on \emph{inclusion through design}, for example, by adopting flexible communication norms, neuro-inclusive hiring practices, and supportive management training. As prior work emphasizes, neurodiversity-aware environments can enhance innovation and problem-solving by embracing diverse cognitive perspectives~\cite{sarker2025inclusive}. 

\subsection{Stress and Burnout}
Resigning from the job is one of the most mentioned negative impacts perceived in \textbf{Q3}:

\interviewquote{Reporting bullying to HR is a waste of time, just quit as fast as you can}

However, most of the resignation comments were not due to any particular discrimination factor but rather to a cumulative effect:
\interviewquote{Dealing with workplace issues takes 90\% of my effort. I evidently get paid to get kicked around, not to produce software.}

\interviewquote{Corporations a legal team on staff; most of us can't afford to hire a lawyer; quitting is the only path forward.}

People who stayed mentioned burnout, depression or an option for seeking therapy:

\interviewquote{I developed a depression because it felt so useless what I did in my job.}

\interviewquote{It would have been nice to have another option: “dealt with stress in a destructive (albeit legal) manner (e.g. alcohol etc)”,}

\subsection{Company Policies and Unintended Consequences}


A recurring theme in our analysis was frustration with the way organizational diversity and inclusion (D\&I) initiatives are implemented. Some participants described supportive and inclusive workplaces, while others criticized policies that inadvertently promoted division or reverse discrimination.  
In several responses, participants expressed discomfort with perceived ideological enforcement or tokenism, echoing earlier concerns that D\&I programs must balance representation with fairness and transparency~\cite{codefirstgirls2024neuro}. Misguided implementations, such as hiring quotas without cultural transformation, can erode trust and reinforce polarization.  

These findings suggest that companies must move beyond surface-level interventions toward evidence-based strategies that foster belonging for all employees. Initiatives should be grounded in continuous feedback, equitable processes, and recognition of intersectional identities rather than single-axis diversity metrics.

\subsection{Never Felt Discrimination: Positive Signals and Blind Spots}
A significant subset of participants reported that they never experienced discrimination personally.
\interviewquote{Despite being trans and non-binary, I have never experienced any workplace discrimination.}

However, many of these respondents, most commonly self-identified as straight, cisgender, white men.  
\interviewquote{I am a white hetero cis man, so, I am not concerned by these discrimination}

\interviewquote{White validate cis het male here, can't be discriminated}

Some also mentioned that, although they did not experience discrimination themselves, they have seen it happen.
\interviewquote{I didn't, but I have seen it happen to several colleagues (Race, gender, sexual orientation, religion and caregiving status)}

This duality underscores both progress and persistent blind spots. On the one hand, some groups experience greater privilege and psychological safety, which can reflect positive aspects of certain workplace cultures. On the other hand, the lack of personal exposure to discrimination can obscure systemic inequities and reduce empathy toward affected colleagues.  
A balanced interpretation recognizes that the perception of fairness can co-exist with unrecognized bias. Encouraging allyship, bias literacy, and open dialogue can help bridge these perceptual gaps and strengthen inclusion across teams.

\subsection{Changing Appearances and Identity Management}

One of the most striking findings of this study was the prevalence of self-modification behaviors, developers changing their appearance, tone, or demeanor to fit workplace expectations. Similar trends have been observed in other contexts. 
Appearing young, which captures controversial strategies such as modifying your résumé to disguise age-related aspects, as well as undergoing plastic surgery to look younger, an aspect that has recently been picked up by major US news outlets~\cite{holley2020cosmetictech}. This category further contains adopting patterns of youthful behavior, including working overtime or during weekends, which are strategies known to conflict with other responsibilities such as family. 
For example, \citeauthor{DBLP:journals/software/BaltesPS20} discuss how older developers sometimes conceal their age by altering résumés or adopting ``youthful'' work patterns~\cite{DBLP:journals/software/BaltesPS20}. Likewise, a recent study reports that women and non-binary developers often change hairstyles, dress styles, or voices to gain professional legitimacy~\cite{van2023still} .  

These acts reveal a deeper cultural tension: the expectation to conform to implicit norms of what a ``developer'' should look or sound like. Such pressures, though subtle, can lead to emotional fatigue, identity conflict, and exclusion. Addressing them requires cultural change within software organizations, valuing authenticity over conformity and challenging stereotypes about professionalism and competence.

\subsection{Threats to Validity} 
Several factors may influence the validity of our findings. 
\subsubsection{Self-selection Bias} As participation in the survey was voluntary, individuals with stronger opinions or experiences related to discrimination may have been more likely to respond, potentially skewing the results. Moreover, since all data are self-reported, participants may underreport or overreport their experiences due to recall limitations or social desirability bias. 

\subsubsection{Representativeness}
Although the survey reached a global audience, its distribution primarily through online developer communities may underrepresent certain populations, such as individuals without consistent internet access or developers in regions with lower participation in global surveys. Consequently, the findings may not fully capture the perspectives of all geographic or demographic segments of the software engineering workforce.  
Furthermore, since our study relies on secondary survey data, we had no control over participant recruitment procedures or regional outreach strategies. These factors may have introduced sampling biases that we could only partially assess by comparing with publicly available workforce demographics.

\subsubsection{Question Framing}
Because the survey instrument was designed externally, we had no influence on the wording, structure, or ordering of the discrimination-related items. The phrasing and predefined categories (e.g., age, gender, race) may have shaped how respondents interpreted and reported discrimination experiences, potentially emphasizing some forms of bias while excluding others. Furthermore, linguistic or cultural variations could have affected the way sensitive topics were understood and disclosed.  
To mitigate this, we consulted with the survey organizers to validate our interpretation of key items, cross-referenced open-ended responses for emergent themes beyond predefined categories, and contextualized our findings accordingly.

\subsubsection{Internal Validity} Given the scale of the dataset (over 8{,}000 survey responses), it is possible that some relevant comments or questions were inadvertently excluded from our analysis. Large-scale survey data often contain substantial noise, including off-topic or ambiguous responses that can complicate systematic inclusion. To mitigate this threat, we focused our qualitative analysis on a subset of 800 open-ended responses, reaching thematic saturation across this sample. Furthermore, since our research specifically targets discrimination-related experiences, we systematically included all responses tagged with the keyword ``discrimination'' in the survey dataset. 

\section{Conclusion}
\label{sec:conclusion}

Our analysis of the \emph{State of Devs 2025} survey reveals that discrimination in software development is multifaceted, extending beyond gender and race to include age, neurodiversity, political beliefs, and caregiving responsibilities. Subtle systemic biases affect developers' well-being, career growth, and sense of belonging, with some modifying behavior to avoid discrimination. At the same time, many participants who belong to a specific group did not report having personal experience of discrimination. 
We hope that our study raises awareness in the research community that discrimination in software development is multifaceted.
We want to encourage researchers to select and assess relevant facets beyond age and gender when designing software engineering studies.

\section*{Acknowledgment}

We thank the organizers of the \emph{State of Devs 2025} survey for providing access to the dataset and making their API publicly available. We are also grateful to the participants who shared their experiences.  
%


\balance
\bibliographystyle{ACM-Reference-Format}
\bibliography{references}

\end{document}